\documentstyle[aps,prd,twocolumn,epsf]{revtex}

\begin{document} 

\newcommand{\sumint}[1]
{\;
{\textstyle\sum}\hspace{-1.1em}
{\displaystyle\int\limits_{#1}}
\;} 

\newcommand{\ket}[1]{\left|#1\right>} 
\newcommand{\bra}[1]{\left<#1\right|} 
\newcommand{\braket}[2]
{\left<#1|#2\right>} 

\newcommand{\ner}[1]{\left|#1\right)} 
\newcommand{\inn}[1]{\left(#1\right|} 
\newcommand{\inner}[2]
{\left(#1|#2\right)}

\title{On the Hawking effect}

\author{Ralf Sch\"utzhold\\
Institut f\"ur Theoretische Physik, 
Technische Universit\"at Dresden\\
D-01062 Dresden, Germany\\
Electronic address : {\tt schuetz@theory.phy.tu-dresden.de}} 

\date{\today}

\maketitle

\begin{abstract} 
In terms of the Painlev{\'e}-Gullstrand-Lema{\^\i}tre coordinates 
a rather general scenario for the gravitational collapse of an object 
and the subsequent formation of a horizon
is described by a manifestly $C^\infty$-metric. 
For a 1+1 dimensional model of the collapse the leading
contributions to the Bogoliubov coefficients are calculated
explicitely and the Hawking temperature is recovered.  
But depending on the particular dynamics of the collapse the final
state represents either evaporation or anti-evaporation. 
The generalization of the calculation to 3+1 dimensions is outlined
and possible implications are addressed.
\end{abstract} 
 
\bigskip 
 
PACS-numbers:  
04.70.Dy,
04.70.-s,
04.62.+v. 


\section{Introduction}\label{introduction}

One of the great challenges of theoretical physics is the quest for  
an underlying law that unifies quantum theory and general relativity. 
The investigation of quantum fields in curved space-times is expected
to provide a chance of achieving some progress towards this aim.    
Fulling's discovery \cite{fulling} of the non-uniqueness of the particle 
interpretation in curved space-times may be regarded as a basis for
various fundamental effects, see 
e.g.~\cite{fulling,zeldovich,eardley,hawking,boulware,unruh,hartle,israel,wald,davies+fulling,fulling+sweeny,rama,haag,kay,dimock,hiscock,fredenhagen,kay+wald,strominger,brout,bousso,nojiri,verch,blackholes,birrell,fulling-buch,visser} 
and references therein.
Perhaps the most prominent example is the Hawking effect \cite{hawking}
which predicts the evaporation of black holes.
There are two alternatives for the investigation of this striking effect:
Originally Hawking calculated the Bogoliubov coefficients via the
geometric optics approximation (backwards ray-tracing) in
Ref.~\cite{hawking}. In contrast to this dynamical treatment Unruh
\cite{unruh} imposed boundary conditions on the state in the static
{\em r\'egime} in order to reproduce the main features of the Hawking
effect.    
Since the state defined in this way -- the Unruh state --  
is completely stationary, it merely describes the late-time part of
the radiation.  
Of course, in general there exists some temporary amount of created
particles that depends on the dynamics of the collapse.   
But according to Ref.~\cite{hawking} the number of these particles
is finite with the result that they disperse after a finite
period of time and thus do not affect the (divergent) late-time
radiation. 
  
Ergo it appears quite natural to assume that the state after the
complete gravitational collapse coincides up to a finite number of
particles with the Unruh state describing evaporation --
independently of the particular dynamics of the collapse. 
The question of whether this assertion is strictly correct will be
subject of the present article. 
For that purpose we shall calculate the number of created particles
explicitely without employing the geometric optics approximation.
It will turn out that the above statement is justified only for one
particular branch of the rather general class of dynamics of the
collapse. 

This paper is organized as follows: 
In Section \ref{formalism} we set up the basic properties of the
quantum field under consideration. A brief introduction into the
concept of Hadamard states is presented in Sec.~\ref{hadamard}.
The number of created particles is calculated in Section \ref{particle}. 
In Secs.~\ref{schwarzschild} and \ref{PGL} we deduce the 
{{\em eigen\,}}modes in 
terms of the Schwarzschild and the Painlev{\'e}-Gullstrand-Lema{\^\i}tre
coordinates, respectively. 
The Bogoliubov coefficients are derived explicitely in
Secs.~\ref{eikonal-ansatz} and \ref{bogoliubov}.
In Section \ref{energy} the relevant expectation values of the
energy-momentum tensor are calculated. 
We shall close with a summary, some conclusions, a discussion, and an 
outlook. 
 
Throughout this article natural units with $G=\hbar=c=k_{\rm B}=1$ 
will be used.
Lowercase Greek indices such as $\mu,\nu$ vary from 0 (time) to
3 (space) and describe space-time components
(Einstein sum convention).
Uppercase Roman indices $I,J$ denote complete sets of quantum numbers. 

\section{General formalism}\label{formalism}

We consider a minimally coupled, massless and neutral (i.e.~Hermitian)
scalar (spin-zero) quantum field $\hat\Phi$ propagating on a globally
hyperbolic space-time $({\cal M},g^{\mu\nu})$. Global hyperbolicity demands
strong causality and completeness, cf.~\cite{ellis}.
(Without these requirements the time-evolution of the quantum system is
not well-defined and unitary.) In the Heisenberg representation the
kinematics of the field $\hat\Phi$ is governed by the
Klein-Fock-Gordon equation
\begin{eqnarray}
\label{KFG}
\Box\hat\Phi=\frac{1}{\sqrt{-g}}\,
\partial_\mu\left(\sqrt{-g}\,g^{\mu\nu}\partial_\nu\hat\Phi\right)=0
\,.
\end{eqnarray}
Strictly speaking, the quantum field is represented by an
operator-valued distribution $\hat\Phi$ and hence the above equation
has to be understood in this sense: 
$\Box F=0\rightarrow\hat\Phi[F]=0$. In a globally hyperbolic
space-time the wave equation (\ref{KFG}) possesses unique advanced
and retarded Green functions 
$\Delta_{\rm adv}(\underline x,\underline x')$ and   
$\Delta_{\rm ret}(\underline x,\underline x')$, respectively. 
Employing these distributions one may accomplish the canonical
quantization procedure via imposing the covariant commutation
relations  
\begin{eqnarray}
\label{green}
\left[\hat\Phi(\underline x),\hat\Phi(\underline x')\right]=
\Delta_{\rm ret}(\underline x,\underline x')-
\Delta_{\rm adv}(\underline x,\underline x')
\,.
\end{eqnarray}
The solutions of the equation of motion (\ref{KFG}) obey a 
symplectic structure induced by the inner product
\begin{eqnarray}
\label{innerP}
\inner{F}{G}=i\int\limits_\Sigma d\Sigma^\mu  
\;F^*\stackrel{\leftrightarrow}{\partial}_\mu G
\,,
\end{eqnarray}
with
$F\stackrel{\leftrightarrow}{\partial}_{\mu}\,  
G=F\,\partial_{\mu}\,G-G\,\partial_{\mu}\,F$.
With the aid of Gauss' law one can show that the inner product
(\ref{innerP}) is independent of the particular Cauchy surface
$\Sigma$ for any two solutions of the Klein-Fock-Gordon equation 
$\Box F=\Box G=0$, cf.~\cite{ellis}.  
It should be mentioned here that the measure $d\Sigma^\mu$ used above
already contains volume factors like $\sqrt{-g_{\Sigma}}$ and is
normalized according to $d\Sigma_\mu\,dx^\mu=\sqrt{-g}\,d^4x$. 

The canonical commutation relations (\ref{green}) imply 
\begin{eqnarray}
\left[\inner{F}{\hat\Phi},\inner{\hat\Phi}{G}\right]=\inner{F}{G}
\,.
\end{eqnarray}
As a result the inner product of the field $\hat\Phi$ with positive
($F_I$) and negative ($F_I^*$) frequency solutions of the
Klein-Fock-Gordon equation, respectively, with
$\inner{F_I}{F_J}=-\inner{F_I^*}{F_J^*}=\delta(I,J)$ and 
$\inner{F_I}{F_J^*}=0$ defines creation and annihilation operators,
respectively. As it is well-known, these operators and thus also the
associated number operators depend on the particular choice of the
solutions $F_I$. This ambiguity represents the non-uniqueness of the
particle interpretation (see e.g.~\cite{fulling}) and may be regarded
as the basis of the phenomenon of particle creation induced by the
gravitational field.    

Averaging the operator-valued distributions $\hat\Phi(\underline x)$
with $n$-point test functions $B_n\in C_0^\infty({\cal M}^n)$ via
\begin{eqnarray}
\hat\Phi^n[B_n]
&=&
\int d^4x_1\cdots\int d^4x_n\,
\hat\Phi(\underline x_1)\cdots\hat\Phi(\underline x_n)
\nonumber\\
&&\times
B_n(\underline x_1,\dots,\underline x_n)
\end{eqnarray}
we acquire well-defined operators $\hat\Phi^n[B_n]$.
The complete set of all these operators (constructed for all test
functions) generates the $*$-algebra containing all possible
observables of the quantum system 
(with the unit element ${\bf 1}=\hat\Phi^0[1]$).

The states $\varrho$ of the quantum system can be introduced as linear 
$\varrho(\mu\hat X+\nu\hat Y)=\mu\varrho(\hat X)+\nu\varrho(\hat Y)$
and non-negative $\varrho(\hat Z^\dagger \hat Z)\ge0$ functionals over the 
$*$-algebra with unit norm $\varrho({\bf 1})=1$.
All these states $\varrho$ build up a convex set, i.e., for any two states
$\varrho_1$ and $\varrho_2$ also the convex combination
$\varrho_\lambda=\lambda\varrho_1+(1-\lambda)\varrho_2$ with
$0<\lambda<1$ represents an allowed state. 
The extremal points of this convex set correspond to the pure states
$\hat\varrho=\ket{\Psi}\bra{\Psi}$.
Since every convex set is the convex hull of its extremal points all 
(mixed) states can be written as a (possibly infinite) linear
combination of pure states. 

In order to decide whether a state is pure or mixed in character one
has to consider the complete algebra. Focusing on a sub-algebra a pure
state may display properties that are usually connected with mixed
states. This observation might be regarded as the basis of the
thermo-field formalism, see e.g.~\cite{thermo-field} and \cite{israel}.

It might be interesting to illustrate these points by some examples:
The space-time obeying the Schwarzschild geometry possesses a Killing
vector mediating the associated (Schwarzschild) time-evolution. 
The ground state of the quantum field (with respect to that Killing
vector) in the region outside the horizon is called the Boulware
\cite{boulware} state $\varrho_{\rm B}$. It contains no particles --
again with respect to the Killing vector measuring the time of an
outside observer with a large and fixed spatial distance to the center
of gravity. 
(A free-falling explorer may well detect particles in that state.)
Ergo the Boulware state is a pure state with respect to the algebra
of the exterior region 
$\hat\varrho_{\rm B}=\ket{\Psi_{\rm B}}\bra{\Psi_{\rm B}}$. 
The interior domain possesses no ground state 
at all, cf.~\cite{blackholes}. (Again all assertions refer to the
Killing field along the Schwarzschild time.)  
As further interesting states one may introduce the
Kubo-Martin-Schwinger (KMS, \cite{kubo}) states $\varrho_{\rm T}$
describing thermal equilibrium at some given temperature $T$.  
Obviously these states are mixed in character -- at least from the
exterior point of view. 
One important example is the
Israel-Hartle-Hawking \cite{hartle,israel} state $\varrho_{\rm IHH}$
which satisfies (in the exterior region) the KMS condition 
corresponding to the Hawking temperature.  
It can be shown \cite{kay} that this state is indeed a pure state
with respect to an enlarged algebra. 
The Israel-Hartle-Hawking state $\varrho_{\rm IHH}$ contains the same
number of ingoing and outgoing particles (thermal equilibrium). Hence
the total energy flux vanishes. 
The phenomenon of the black hole evaporation can be described by the
Unruh \cite{unruh} state $\varrho_{\rm U}$. This state is defined via
two requirements: no ingoing/incoming particles/radiation at spatial
infinity and thermal outgoing radiation near the horizon, see also
\cite{dimock}.    
If one considers a gravitational collapse of an object and assumes the
initial state to be pure in character (e.g.~the vacuum) then the final
state is -- of course -- also a pure state.
(Here the notion of a pure state refers in both cases to the complete
algebra.)  
The question of whether the initial state indeed transforms into the
Unruh state will be subject of Section \ref{particle}.       

\subsection{Hadamard states}\label{hadamard}

In general, the complete convex set is too large and contains more
states than physical reasonable. One way to restrict to physically
well-behaving states is to impose the so-called Hadamard
\cite{hadamard} condition. Hadamard states are states for which the
symmetric part of the bi-distribution 
\begin{eqnarray}
{\cal W}^{(2)}(\underline x,\underline x')=
\varrho\left(\hat\Phi(\underline x)\hat\Phi(\underline x')\right)
={\rm Tr}\left\{
\hat\varrho\,\hat\Phi(\underline x)\hat\Phi(\underline x')
\right\}
\,,
\end{eqnarray}
the two-point Wightman \cite{wightman} function, 
obeys the following singularity structure
(in a 3+1 dimensional space-time)
\begin{eqnarray}
\label{hadi}
&\phantom{x}&
\frac{1}{2}
\left({\cal W}^{(2)}(\underline x,\underline x')+
{\cal W}^{(2)}(\underline x',\underline x)\right)= 
\nonumber\\
&\phantom{x}&
-\frac{1}{(2\pi)^2}{\cal P}\left(
\frac{U(\underline x,\underline x')}{s^2}+
V(\underline x,\underline x')\ln s^2+
W(\underline x,\underline x')
\right)
\,,
\end{eqnarray}
where $\cal P$ symbolizes the principal part.
The antisymmetric part of ${\cal W}^{(2)}$ must be consistent with the
commutation relation (\ref{green}). $s$ denotes the geodesic distance
$ds^2=g_{\mu\nu}dx^\mu dx^\nu$ between the space-time points 
$\underline x$ and $\underline x'$ 
(which is at least in a neighborhood of a regular point 
$\underline x$ unique).
The functions $U(\underline x,\underline x')$, 
$V(\underline x,\underline x')$ and $W(\underline x,\underline x')$ 
are regular in the coincidence limit $\underline x\rightarrow\underline x'$.
Together with the normalization $U(\underline x,\underline x)=1$ the
first two functions  
$U(\underline x,\underline x')$ and $V(\underline x,\underline x')$
are uniquely determined by the  
structure of space-time, e.g.~$V(\underline x,\underline x)=R^\mu_\mu/12$ 
(with $R_{\mu\nu}$ being the Ricci tensor, see e.g.~\cite{hadamard}).
Hence all information about the state $\varrho$ enters 
$W(\underline x,\underline x')$ only.   
 
One important advantage of the Hadamard requirement may be illustrated
by considering the regularization of expectation values of two-field
observables, for instance the energy-momentum tensor 
$\hat T_{\mu\nu}$. The Hadamard singularity structure ensures the
validity of the point-splitting  renormalization technique,
cf.~\cite{wald}.

It can be shown that for a globally hyperbolic $C^\infty$ 
(i.e.~infinitely differentiable) space-time 
$({\cal M},g_{\mu\nu})$ the Hadamard condition is conserved, i.e., 
if the two-point function has the Hadamard singularity structure in an
open neighborhood of a Cauchy surface, then it does so everywhere
\cite{fulling+sweeny}. 

If one considers the gravitational collapse of an object which can be
described by a $C^\infty$-metric the above theorem can be used to
deduce the Hadamard condition for the final state. (The initial state
is assumed to be a regular excitation over the ground state and thus
satisfies the Hadamard requirement. The Minkowski vacuum of course
meets the Hadamard structure with $U=1$, $V=0$ and $W=0$.)   
On the other hand it can be shown that if the state of a field
$\hat\Phi$ fulfills the Hadamard requirement  
(among further not as strict assumptions, cf.~\cite{fredenhagen}) 
in the whole space-time of the Schwarzschild geometry and especially
at the horizon   
then the asymptotic expectation values correspond exactly to a 
thermal radiation with the Hawking temperature  $T=1/(4\pi R)$  
(see \cite{fredenhagen}, \cite{haag}, and \cite{kay+wald}).
Combining the two statements above we are able to deduce the 
Hawking temperature for any collapse scenario that can be described by
a $C^\infty$-metric. 

It might be interesting to discuss the previous considerations by means
of some examples for the Schwarzschild geometry. 
Applying the theorems above to the Boulware state,
i.e.~the ground state, it follows immediately that this state cannot
satisfy the Hadamard requirement -- at least at the horizon.
Indeed this state is singular at the horizon -- its (point-splitting)
renormalized energy density diverges there 
$\bra{\Psi_{\rm B}}\hat T^0_0\ket{\Psi_{\rm B}}_{\rm ren}
\downarrow-\infty$ for $r\downarrow R$. 
It can be shown that the Boulware state as well as every KMS state
(with an arbitrary temperature) fulfills the Hadamard requirement {\em away}
from the horizon $r>R$, see \cite{verch}. 
But only the KMS state corresponding to the Hawking temperature,
i.e.~the Israel-Hartle-Hawking state (after a suitable extension)
meets the Hadamard structure {\em at} the
horizon, see e.g.~\cite{verch,kay+wald}. 
However, the initial (approximately Minkowski) vacuum cannot transform
into this state during a gravitational collapse of an object,
cf.~\cite{blackholes}. In contrast to the Unruh state the
Israel-Hartle-Hawking state represents 
thermal equilibrium also for $r\uparrow\infty$ and the associated
amount of particles and energy cannot be produced by a collapse,
see also Secs.~\ref{bogoliubov} and \ref{discussion} below.
     
\section{Particle creation}\label{particle}

Within the Heisenberg representation the time-evolution of the quantum
system is governed by the operators while the states remain
unaffected. Hence the investigation of the Hawking effect goes along 
with the question: How many (final) Schwarzschild particles contains the
initial state? In general, this number depends on the particular
initial state and the initial metric as well as the dynamics of the
metric during the collapse. According to the considerations in the
previous Section we assume a $C^\infty$-metric throughout. It can be
shown that the Hawking effect (i.e.~the late-time radiation) is
independent of the (regular) initial space-time, 
see Sec.~\ref{bogoliubov} below. Similarly any finite amount of
particles being present initially does not alter the assertions
concerning the Hawking effect (see the remarks at the end of Section
\ref{bogoliubov} below). 
For that reason we assume the initial state to coincide with the
(initial) vacuum. In this situation the number of final particles can
be calculated via the Bogoliubov $\beta$-coefficients, 
see e.g.~\cite{birrell}       
\begin{eqnarray}
\label{w-int}
N_J^{\rm out}=\bra{0^{\rm in}}\hat N_J^{\rm out}\ket{0^{\rm in}}=
\sumint{I}|\beta_{IJ}|^2
\,.
\end{eqnarray}
In order to calculate these coefficients we have to derive the structure
of the initial modes $F_I^{\rm in}$ after the collapse and to compare
them with the out-solutions $F_J^{\rm out}$ by means of the inner
product in Eq.~(\ref{innerP}).

\subsection{Schwarzschild metric}\label{schwarzschild}

The particle interpretation in quantum field theory is based on
the selection of an appropriate time-like Killing vector.
This choice refers to a certain class of associated observers
whose time-evolution is generated by the Killing field.
For the flat space-time example, the Killing vector mediating the
(Minkowski) time translation symmetry accords to a usual beholder at
rest whereas special Lorentz boosts represent accelerated (Rindler)
explorers. 
Since, in general, different Killing vectors generate distinct
particle definitions, the Rindler observer does not regard
the Minkowski vacuum as empty with respect to (Rindler) particles.  
Instead, he/she experiences a thermal bath, 
a phenomenon which is called the Unruh effect \cite{unruh}.

The time-evolution of a beholder at a large and fixed spatial
distance to the center of gravity is generated
by the Killing vector corresponding to the Schwarzschild time $t$. 
The particles that are measured by such an observer can be described
by positive frequency solutions -- with respect to that time coordinate
-- of the Klein-Fock-Gordon equation. 
In contrast the evolution parameters of further coordinate
representations of the Schwarzschild geometry 
(e.g.~the Lema{\^\i}tre metric) represent
different explorers (e.g.~the free-falling one) in general.

In terms of the Schwarzschild coordinates $t,r,\vartheta,\varphi$ the
3+1 dimensional metric assumes the well-known form
\begin{eqnarray}
\label{3+1}
ds^2
&=&
\left(1-\frac{R}{r}\right)dt^2-\left(1-\frac{R}{r}\right)^{-1}dr^2
\nonumber\\
\nonumber\\
&&
-r^2\,d\vartheta^2-r^2\,\sin^2\vartheta\,d\varphi^2             
\,,
\end{eqnarray}
where $R$ denotes the Schwarzschild radius and describes the position
of the horizon.

Strictly speaking, there exist several definitions of a horizon, for 
example the event, the apparent, and the putative horizon,
cf.~\cite{ellis} and \cite{visser}.
The notion of the event horizon refers to the global structure of the 
space-time (asymptotical reachability) whereas the apparent horizon can
be defined by strictly local considerations (trapped surfaces). 
Together with some additional requirements 
(e.g.~asymptotical flatness, cf.~\cite{visser}) 
also the putative horizon represents a local condition: 
'time slows to a stop', cf.~\cite{visser}.
In the space-time of the eternal Schwarzschild geometry 
(see Figs.~\ref{fig1} and \ref{fig2})
all these
definitions coincide, but in a dynamical space-time describing the
gravitational collapse of an object 
(see Figs.~\ref{fig4} and \ref{fig5})
this coincidence does not hold in general.
Within our investigations we always refer to a locally defined horizon
-- such as the apparent horizon. 

As it will become more evident later on, the most interesting region
(with respect to the Hawking effect) is the vicinity of the horizon.
In order to extract the features that are characteristic for this zone
we introduce a dimensionless variable $\chi$ via
\begin{eqnarray}
\chi=\frac{r}{R}-1
\,.
\end{eqnarray}
This quantity allows for a Taylor expansion in the vicinity of the
horizon. As another useful tool we define the Regge-Wheeler tortoise
coordinate
\begin{eqnarray}
r_*=\int\frac{dr}{1-R/r}=R\ln\chi+{\cal O}[\chi]
\,.
\end{eqnarray}
For reasons of simplicity we restrict our further considerations to the
$(t,r)$-sector and drop the angular contributions in Eq.~(\ref{3+1}). 
The resulting 1+1 dimensional space-time obeys a conformally flat
metric when expressed in terms of the tortoise coordinate 
\begin{eqnarray}
ds^2=\left(1-\frac{R}{r}\right)\left(dt^2-dr^2_*\right)
\,.
\end{eqnarray}
As a result the equation of motion (\ref{KFG}) simplifies to
\begin{eqnarray}
\left(
\frac{\partial^2}{\partial t^2}-
\frac{\partial^2}{\partial r_*^2} 
\right)\Phi=0
\end{eqnarray}
in 1+1 dimensions.
(In 3+1 dimensions additional terms occur and generate slight
modifications, see Sec.~\ref{outlook}.)
As a result the positive frequency Schwarzschild 
{{\em eigen\,}}functions are given by  
\begin{eqnarray}
\label{eigenBH}
F_I^{\rm out}(\underline x)
&=&
{\cal N}^{\rm out}\,
\frac{e^{-i\omega t\pm i\omega r_*}}{\sqrt{\omega}}
=
F_{\xi\omega}^{\rm out}(t,\chi)
\nonumber\\
&=&
{\cal N}^{\rm out}\,
\frac{e^{-i\omega t}}{\sqrt{\omega}}\,
\chi^{-i\xi\omega R}\,
\left(1+{\cal O}[\chi]\right)
\end{eqnarray}
for $r>R$ and vanish for $r<R$ due to the horizon,
cf.~\cite{blackholes}.  
The ingoing and outgoing modes are distinguished by $\xi=\pm1$.
(For the modifications occurring in 3+1 dimensions see
Sec.~\ref{outlook}.) 
${\cal N}^{\rm out}$ symbolizes a normalization factor
which may without any loss of generality chosen to be independent of  
$\xi$. 
These {{\em eigen\,}}functions are rapidly oscillating near the horizon 
which again hints that there is the most interesting region:
This singular behavior of the modes corresponds to the freezing of
the kinematics of the field (governed by the Klein-Fock-Gordon
equation) in the vicinity of the (putative) horizon.

\subsection{Painlev{\'e}-Gullstrand-Lema{\^\i}tre metric}\label{PGL} 

The Schwarzschild metric is quite simple but exhibits a coordinate
singularity at the horizon and is therefore not $C^\infty$ there.
Hence it is impossible to express a manifestly $C^\infty$-metric in
terms of the Schwarzschild coordinates. For this purpose one has to
employ other coordinate systems. As one possible candidate we consider
the Painlev{\'e}-Gullstrand-Lema{\^\i}tre \cite{PaGuLe} coordinates
$t_{\rm PGL},r,\vartheta,\varphi$.
These coordinates emerge from the Schwarzschild coordinates 
$t_{\rm S},r,\vartheta,\varphi$ by means of the transformation
\begin{eqnarray}
dt_{\rm PGL}=dt_{\rm S}+\sigma\frac{\sqrt{R/r}}{1-R/r}dr
\,.
\end{eqnarray}
There exist two branches of these coordinate set distinguished by
$\sigma=\pm1$. In the following we shall drop the index 
$t=t_{\rm PGL}$ for convenience. The metric transforms into
\begin{eqnarray}
\label{statPGL}
ds^2
&=&
\left(1-\frac{R}{r}\right)dt^2-2\sigma\sqrt{\frac{R}{r}}\,drdt-dr^2 
\nonumber\\
\nonumber\\
&&
-r^2\,d\vartheta^2-r^2\,\sin^2\vartheta\,d\varphi^2
\,.
\end{eqnarray}
\begin{figure}[ht]
\centerline{\mbox{\epsfxsize=8cm\epsffile{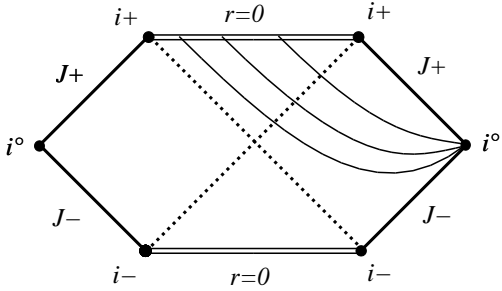}}}
\caption{
Penrose diagram of the maximally extended Kruskal manifold.
Owing to the conformal mapping the null lines (light rays) are at 
$\pm45^\circ$.
The horizontal axis of symmetry indicates the borderline between
future (above) and past (below).
Adopting the notation of Hawking and Ellis the future and the past
infinity $J+$ and $J-$ (with $t=+\infty$ and $t=-\infty$, respectively,
as well as $r=\infty$) are denoted by single solid lines. 
Double solid lines symbolize the future and past singularity at $r=0$. 
The horizons at $r=2M$ are represented by dotted lines.
$i^\circ$ denotes the spatial infinity (with $r=\infty$ and $t$ finite).
$i+$ and $i-$ symbolize the future and the past, respectively, (with
$t=+\infty$ and $t=-\infty$, respectively, and $r$ finite).
Representative surfaces of constant PGL time are indicated for
$\sigma=+1$. 
This branch of the coordinates merely covers the black hole (future) 
horizon and singularity -- but not the white hole (past) horizon or
singularity.}  
\label{fig1}
\end{figure}

\begin{figure}[ht]
\centerline{\mbox{\epsfxsize=8cm\epsffile{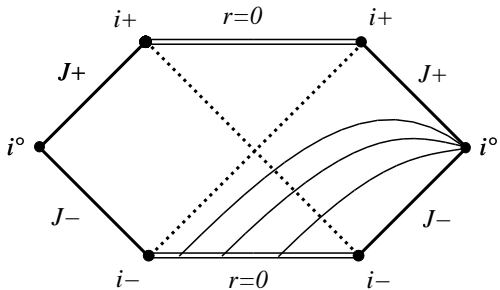}}}
\caption{
Penrose diagram of the Kruskal manifold with representative surfaces
of constant PGL time for the branch $\sigma=-1$. 
Obviously this figure can be obtained by time-reversing
the previous diagram.
The $\sigma=-1$ branch of the PGL coordinates covers the white hole 
(past) horizon and singularity.}
\label{fig2}
\end{figure}

Although the Painlev{\'e}-Gullstrand-Lema{\^\i}tre (PGL) metric does not
belong to the well-known and frequently discussed representations of
the Schwarzschild geometry 
(such as the Kruskal, Eddington-Finkelstein, Novikov, or Lema{\^\i}tre
coordinates) it possesses several advantages:

In contrast to the Schwarzschild form the PGL metric 
(and its inverse as well)  
is $C^\infty$ except at the singularity at $r=0$.\\  
The hyper-surfaces of constant PGL time $dt=0$ are equivalent to flat
Euclidean spaces; the space-time curvature is encoded in the shift
vector (employing the Arnowitt-Deser-Misner (ADM) notation).\\
The evolution parameter, i.e.~the PGL time, corresponds to a Killing
vector leading to a stationary metric.
This fact simplifies the particle definition via positive frequency 
solutions.\\  
Asymptotically $r\uparrow\infty$ the PGL representation coincides with
the Minkowski metric (similar to the Schwarzschild form).
By virtue of Birkhoff's theorem this coincidence persists during the
dynamical period of the collapse.\\
The radial coordinate $r$ directly corresponds to the surface of the
two-sphere $\{t,r={\rm const}\}$ via $4\pi r^2$.\\   
Last but not least the effective acoustic metric of the sonic
analogues \cite{sonic} of the Schwarzschild geometry equals 
-- up to a conformal factor -- the PGL form, see e.g.~\cite{acoustic}.  

For further discussions of the properties of the PGL metric see 
e.g.~Refs.~\cite{kraus,martel,israel-buch,sonic,acoustic,volovik,fischer} 
and references therein.
For example Ref.~\cite{martel} presents a pedagogical presentation of
the PGL metric as well as its relation to the Eddington-Finkelstein
coordinates. 

It should be mentioned here that the PGL coordinates do not cover the
complete fully extended Kruskal manifold:
E.g., depending on the particular branch (i.e.~the sign of $\sigma$)
the PGL representation covers either the future (black hole) event
horizon and the future (black hole) singularity for $\sigma=+1$ or the
past (white hole) event horizon and the past (white hole) singularity
for $\sigma=-1$, see Figs.~\ref{fig1} and \ref{fig2}.  
However, both branches possess an apparent horizon at $r=R=2M$. 

Since we shall calculate the inner product in terms of the new
coordinates we have to transform the Schwarzschild {{\em eigen\,}}functions,
i.e.~the $\rm out$-modes. This can be done by simply substituting the
Schwarzschild time via 
$t_{\rm S}=t_{\rm PGL}-\sigma R \ln\chi+{\cal O}[\chi]$
in Eq.~(\ref{eigenBH})
\begin{eqnarray}
\label{eigenPGL}
F_I^{\rm out}(\underline x)
=
{\cal N}^{\rm out}\,
\frac{e^{-i\omega t}}{\sqrt{\omega}}\,
\chi^{i(\sigma-\xi)\omega R}
\left(1+{\cal O}[\chi]\right)
\,.
\end{eqnarray}
(Again we restrict our considerations to 1+1 dimensions.)
One observes that the modes with $\xi=\sigma$ are no longer singular
(arbitrarily fast oscillating) at the horizon, only those with
$\xi=-\sigma$ still exhibit this property. As it will become evident
later on, merely the singular modes with $\xi=-\sigma$ will contribute to
the Hawking effect.     

Employing the Painlev{\'e}-Gullstrand-Lema{\^\i}tre coordinates it is
possible to write down a manifestly $C^\infty$-metric modeling a
gravitational collapse of an object and the subsequent formation of a
horizon   
\begin{eqnarray}
\label{dynPGL}
ds^2=\left(1-f^2(t,r)\right)dt^2-2\sigma f(t,r)drdt-dr^2 
\,,
\end{eqnarray}
with $f\in C^\infty$. Initially $t\downarrow-\infty$
the metric describes a regular object with a
(relatively) dilute distribution of matter and can be approximated
(locally) by the Minkowski metric 
$f(t\downarrow-\infty,r)=f_{\rm in}(r)\ll1$.
For reasons of simplicity we assume the horizon to be formed at $t=0$,
i.e.~$f(t\geq 0,r\geq R)=f_{\rm out}(r)=\sqrt{R/r}$.
(Note, that we did not impose any conditions on the structure of $f$ in
the interior region, i.e.~beyond the horizon.) 
Outside the (spherically symmetric) collapsing object the Birkhoff
theorem demands a stationary metric $f(t,r \gg R)=\sqrt{R/r}$.

The Jacobi determinant is simply given by $\sqrt{-g}=1$
and the metric as well as its inverse are smooth  
$g_{\mu\nu},g^{\mu\nu}\in C^\infty$. 
Of course, this assertion holds true only if we omit the formation of
the singularity at $r=0$. But the region beyond the horizon is
causally separated from the outside domain and 
(as it will turn out later) irrelevant for our purposes.

Considering the sonic analogues of the Schwarzschild geometry the
function $f(t,r)$ directly corresponds to the time-dependent local
velocity of the fluid, cf.~\cite{sonic,acoustic,volovik,fischer}.

\begin{figure}[ht]
\centerline{\mbox{\epsfxsize=4cm\epsffile{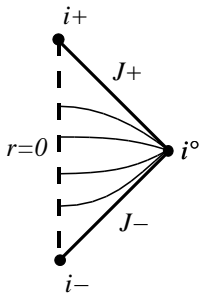}}}
\caption{Penrose diagram of the Minkowski space-time.
The dashed line symbolizes the (regular) origin $r=0$.
Light rays originating from $J-$ bounce off at the origin and
propagate to $J+$. The Minkowski time equals the PGL time for
$f(t,r)=0$. Again representative surfaces of constant (PGL) time are
indicated.}  
\label{fig3}
\end{figure}

\begin{figure}[ht]
\centerline{\mbox{\epsfxsize=5cm\epsffile{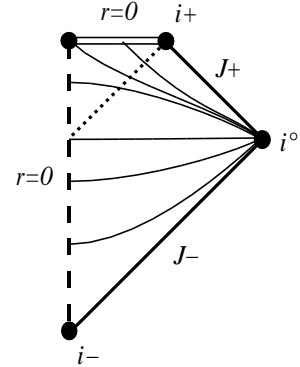}}}
\caption{
Penrose diagram of the collapse to a black hole as described by the
branch $\sigma=+1$ of the PGL metric with an appropriate function
$f(t,r)$. As one can infer from the indicated surfaces of constant PGL
time, the formation of the black hole horizon (dotted line) can be
described regularly by these coordinates (with $\sigma=+1$).} 
\label{fig4}
\end{figure}

\begin{figure}[ht]
\centerline{\mbox{\epsfxsize=8cm\epsffile{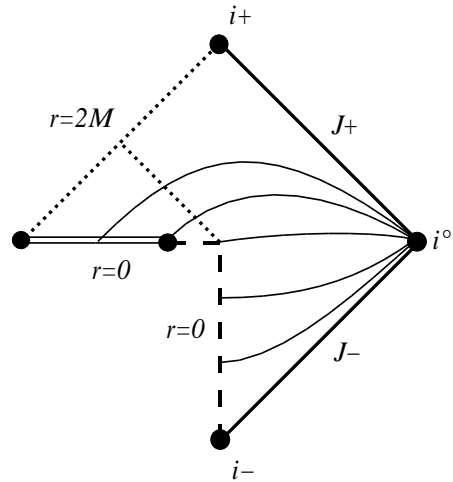}}}
\caption{
Penrose diagram of the collapse to a white hole as described by the
branch $\sigma=-1$ of the PGL metric with an appropriate function
$f(t,r)$.
After the formation of the white hole horizon (dotted line) no light
ray originating from $J-$ can reach $r=0$. 
The particular structure of this figure is based on the (not necessary)
assumption that the singularity at $r=0$ develops at a finite period
of PGL time after the horizon has been formed. 
Again one may infer from the indicated surfaces of constant PGL time
that the formation of the white hole horizon can be described
regularly by the branch $\sigma=-1$ of the PGL metric.
In contrast to Figs.~1 and 2 this diagram cannot be obtained by
time-reversing the previous figure.}
\label{fig5}
\end{figure}

It should be mentioned here that the knowledge of the above metric 
over a finite period of time is not sufficient for determining an 
event horizon -- in contrast to the eternal (stationary) metric in
Eq.~(\ref{statPGL}). 
The local metric above does also not allow for the construction of a
Penrose diagram (see Figs.~\ref{fig1}-\ref{fig5}) -- this requires the
extension to the complete space-time.
Similarly it does not necessarily contain a space-time singularity.
However, one may deduce the existence of an apparent horizon at $r=R$
for $t\geq0$. 

In a purely 1+1 dimensional consideration the range of the coordinate
$r$ in Eq.~(\ref{dynPGL}) might be chosen arbitrarily. But in order to
keep contact to the 3+1 dimensional situation it should be specified
according to $0\leq r <\infty$.  
The 3+1 dimensional bouncing-off effect (see also Fig.~\ref{fig3})
at the origin $r=0$ can be
simulated in 1+1 dimensions by an appropriate boundary condition. As
already stated in Ref.~\cite{unruh}, Dirichlet or Neumann boundary
conditions or every linear combination of them are suitable, see also
the discussions at the end of the next Section. 

Although the two distinct branches $\sigma=\pm1$ of the stationary
Painlev{\'e}-Gullstrand-Lema{\^\i}tre metric in Eq.~(\ref{statPGL}) 
are related to each other via a simple change of the coordinates or
the time-inversion ${\sf T}:t\rightarrow-t$, 
the distinction between the different collapse dynamics for
$\sigma=+1$ and $\sigma=-1$, respectively, in Eq.~(\ref{dynPGL}) 
cannot be removed by any transformation.
(It is not possible to find a globally integrating factor for the
differential form.)
The two branches correspond to two non-equivalent collapse scenarios
(see Figs.~\ref{fig4} and \ref{fig5})
and -- as we shall see later -- generate completely different final
states of the quantum field.  
Nevertheless, in both cases the initial $t\downarrow-\infty$ metric
describes a regular object whereas the final $t\uparrow\infty$ metric
represents 
for $r>0$ a vacuum solution of Einstein's equations with a central
mass $M=R/2$.
So there exists {\em a priori} no reason to prefer one of the two branches,
see also the remarks in Section \ref{discussion}.

\subsection{Eikonal {\em ansatz}}\label{eikonal-ansatz} 

In order to calculate the Bogoliubov coefficients we have to deduce
some informations about the $\rm in$-modes.
For that reason we adopt the eikonal {\em ansatz} and divide the field
into an amplitude and a phase 
\begin{eqnarray}
\label{eikonal}
F_{\xi,\omega}^{\rm in}(t,r)
&=&
\frac{1}{\sqrt{\omega}}\,A_{\xi}(t,r)
\exp\left\{-i\omega S_\xi(t,r)\right\}
\nonumber\\
&&
\times
\left(1+{\cal O}\left[\frac{1}{\omega}\right]\right)
\,.
\end{eqnarray}
Certainly this {\em ansatz} will be justified for compact space-time
domains (which are not too large) with smooth metrics and high (initial) 
frequencies $\omega$ (see also the remarks at the end of this Section). 
But as it will turn out later, this is exactly the limit that is
relevant for the Hawking effect.   
Inserting the above expression into the Klein-Fock-Gordon equation
(\ref{KFG}) the leading terms in $\omega$ govern the kinematics of the
phase function via 
\begin{eqnarray}
({\partial}_\mu S_\xi) 
\,g^{\mu\nu}\, 
({\partial}_\nu S_\xi)
&=&
0\,, 
\quad\quad{\rm i.e.,}
\nonumber\\
({\partial}_t S_\xi-\sigma f {\partial}_r S_\xi)^2
&=&
({\partial}_r S_\xi)^2
\,.
\end{eqnarray}
This non-linear equation has four separate branches of solutions --
e.g.~for $f=0$ one may identify the positive and negative frequency
solutions on the one hand and the ingoing and outgoing components
labeled by $\xi=\pm1$ on the other hand
\begin{eqnarray}
\label{branch}
{\partial}_t S_\xi-\sigma f {\partial}_r S_\xi
&=&
\xi{\partial}_r S_\xi\,,
\quad\quad{\rm i.e.,}
\nonumber\\
{\partial}_t S_\xi
&=&
(\sigma f+\xi){\partial}_r S_\xi
\,.
\end{eqnarray}
However, these four branches will not necessarily be separated for
arbitrary space-time dependent functions $f(t,r)$.
E.g., if $f(t,r)$ oscillates with a large elongation, a mode which is
initially purely ingoing may turn its direction into outgoing and so
on.   
Nevertheless, if we assume a sufficiently well-behaving dynamics of $f$,  
e.g.~if it transforms directly and smoothly from $f_{\rm in}$ to
$f_{\rm out}$ -- where the relevant time-scales are smaller than the
length scales ($R$) -- the four branches remain separated:
In this case the different branches cannot approach each other close
enough during the time-evolution. 
As a limiting case we may consider a very rapid change 
(sudden approximation) of the metric
$f(t,r) \approx f_{\rm in}(r)\Theta(-t)+f_{\rm out}(r)\Theta(t)$. 
In this situation the final phase
function $S_\xi$ coincides (nearly) with its initial form while its
time-derivative changes according to Eq.~(\ref{branch}).
The sudden approximation does not hold in contrast with the high
frequency limit since we deal with the frequency-independent phase
function $S_\xi$.

In summary, the above assumption of a rapid collapse ensures
the separation of  the four branches, e.g.~if 
${\partial}_r S_\xi$ is positive/negative
initially then it remains positive/negative also after the
collapse. 
The same applies to the time-derivative ${\partial}_t S_\xi$ 
-- as long as $f<1$, i.e.~outside the horizon, see Eq.~(\ref{branch}).   
As a consequence the division of the modes into ingoing
and outgoing -- labeled by $\xi=\pm1$ -- can be used throughout. 
 
Additional complications arise in 3+1 dimensions, but the main result
-- the separation of the four branches -- persists under appropriate
assumptions:\\  
As demonstrated in Ref.~\cite{blackholes}, a regular spherically
symmetric 3+1 dimensional space-time without horizon does not allow
for the definition 
of ingoing and/or outgoing particles. The {{\em eigen\,}}modes are
standing 
waves, i.e.~linear combinations of ingoing and outgoing components
with equal weights. 
So the bouncing-off effect at $r=0$ mixes the ingoing and
outgoing components during static as well as during the dynamical
period.  
In a 1+1 dimensional consideration this "reflection" may
be simulated by an effective boundary condition at $r=0$, cf.~the
remarks below Eq.~(\ref{dynPGL}).  
Selecting appropriate coordinates the point $r=0$ becomes
time-dependent. 
E.g., in terms of length and time-scales associated to an outside
observer the center of the collapsing object goes to infinity 
(asymptotically at a null line) owing to the formation of the
horizon. In terms of these particular coordinates the origin $r=0$
corresponds to an accelerated mirror. 
Ref.~\cite{davies+fulling} presents a derivation of the Hawking effect
based on the moving mirror analogue.\\
In contrast, in terms of the Painlev{\'e}-Gullstrand-Lema{\^\i}tre
coordinates the origin $r=0$ obeys no time-dependence at all. 
Again we assume the collapse to occur fast enough:
The metric is presumed to remain stationary until $t=-R$ 
(the beginning of the collapse) and according to Sec.~\ref{PGL} the
(apparent) horizon at $r=R$ is formed at $t=0$.
Since the Schwarzschild {{\em eigen\,}}functions vanish for $r<R$ it is
sufficient to consider the region outside the horizon to be formed at
$r \geq R$. 
Within this limited space-time domain $\{-R\leq t\leq0,r\geq R\}$ 
the ingoing and outgoing components are indeed effectively
independent:
It takes every ingoing light ray inside this domain at least the
time duration $\Delta t = R$ 
(under the assumptions made for the function $f(t,r)$ above)
to propagate to the origin, to bounce off
(turning its direction into outgoing), and to reach the radius $r=R$
again.    
An analogue assumption was already imposed in Ref.~\cite{unruh}. 
In such a scenario the information about a possible 
"reflection" at $r=0$ cannot influence the relevant region. 
In summary we arrive at the conclusion that 
(under the assumptions made) the four branches in Eq.~(\ref{branch})
are indeed effectively independent also in 3+1 dimensions
(see also the discussion at the end of Sec.~\ref{discussion}).

It should be mentioned here that the above {\em ansatz} is not
equivalent to the quasi-classical 
Jeffreys-Wentzel-Kramers-Brillouin (JWKB)
approximation (expansion into powers of $\hbar$)
-- in spite of some similarities.
It does also not coincide with the geometric optics approximation 
(backwards ray tracing) which was used in Ref.~\cite{hawking}.
Instead the eikonal {\em ansatz} is based on a consequent expansion
into inverse powers of the initial frequency $\omega$. 

\subsection{Bogoliubov coefficients}\label{bogoliubov}

Now we are in the position to calculate the Bogoliubov coefficients
and thereby the number of created particles explicitely.
Unfortunately, it seems to be impossible to find a general solution
for these overlap coefficients. Nevertheless, with an expansion into
powers of the relative distance to the horizon $\chi$ and the inverse
initial frequency $1/\omega$ it is possible to extract the leading
contribution -- the Hawking effect.   
(As it will turn out later, the sub-leading parts merely generate
finite contributions and thus do not affect the late-time radiation.)
Per definition the Hawking radiation is exactly that part of the
radiation which persists at arbitrarily late times (if we neglect the
back-reaction). Hence the number of created particles accounting for
the Hawking effect has to diverge. Any finite amount of particles
would disperse after a finite period of time and cannot generate
late-time radiation. (This is a consequence of the spectral properties
of the wave equation. It possesses a purely continuous spectrum and
thus does only allow for scattering states but no bound states, 
see e.g.~\cite{blackholes} and \cite{fulling-buch}.)
As demonstrated in Ref.~\cite{blackholes}, the divergent number of
particles is necessary for the thermal behavior in an infinite volume.        
In order to isolate the divergent part of the number of created
particles we have to consider the Bogoliubov $\beta$-coefficients  
(see e.g.~\cite{birrell})
\begin{eqnarray}
\beta_{IJ}=i\int\limits_\Sigma d\Sigma^\mu  
\;F_I^{\rm in}
\stackrel{\leftrightarrow}{\partial}_\mu 
F_J^{\rm out}
\,.
\end{eqnarray}
Since the Painlev{\'e}-Gullstrand-Lema{\^\i}tre coordinates are
completely regular the measure $d\Sigma_\mu$ does not contain any
singularities. As we have observed in the previous Sections, the modes 
$F_I^{\rm in}$ and $F_J^{\rm out}$ are bounded. In addition, the
Birkhoff theorem implies that the modes at very large spatial
distances to the collapsing object are not affected by the
collapse. Consequently this region does not contribute to the
$\beta$-coefficients and generates a $\delta(\omega-\omega')$-term for
the $\alpha$, see also \cite{hawking}. In summary we arrive at the
conclusion that all (single) Bogoliubov $\beta$-coefficients are
finite. As a result the divergence of the number of created particles 
$N_J^{\rm out}$ must be traced
back to the summation/integration over the initial quantum numbers
$I=(\xi,\omega)$ in Eq.~(\ref{w-int})
\begin{eqnarray}
\nonumber
N_J^{\rm out}=\sumint{I}|\beta_{IJ}|^2
\,.
\end{eqnarray}
There are two possibilities for a singularity, the IR- and the
UV-divergence of the integration over the initial frequencies
$\omega$. 
In the limit of small frequencies $\omega$ the modes
become space- and time-independent and approach a constant --
unaffected by the Klein-Fock-Gordon equation. 
(Here we regard the IR-singular normalization $1/\sqrt{\omega}$ as
factorized out.)
Ergo in the limiting case $\omega\downarrow0$ the 
in- and out-modes coincide and thus possess a vanishing overlap with
all other modes corresponding to finite frequencies. As a consequence
the $\omega$-integration of the (absolute values squared of the)
Bogoliubov coefficients is IR-save.        

In summary the infinite amount of particles has to be caused by the
UV-divergence  of the integration over the initial
frequencies in consistency with Ref.~\cite{hawking}. 
(The Hawking effect is dominated by large (initial) frequencies
only if one considers a fundamental quantum field theory without 
any kind of dispersion.
Introducing a cut-off, see e.g.~\cite{cut-off}, as an effective
description of some underlying theory the calculations are different.)

Recalling the structure of the initial {{\em eigen\,}}functions in
Eq.~(\ref{eikonal}) we arrive at the conclusion that only
singularities of the out-modes may induce a UV-divergence.
The convolution of regular expressions with the for
$\omega\uparrow\infty$ arbitrarily fast oscillating in-modes yields 
results of order $1/\omega$. 
Ergo the subsequent $\omega$-integration would be UV-save. 
Indeed, the out-modes are not regular
at the horizon -- the region that is naturally relevant for the
Hawking effect. Thus it is adequate to consider the vicinity of the
horizon and the high (initial) frequency limit in order to extract the
Hawking effect. As it will become more evident later, exactly the
leading contributions in $\chi$ and $1/\omega$ are sufficient for the
derivation of the thermal radiation.  

If we choose the Cauchy surface according to 
$\Sigma=\{0 \leq r <\infty , t=0\}$ the
surface element assumes the form  $d\Sigma_\mu=(dr,0)$ and the
$\beta$-coefficients transforms into 
\begin{eqnarray}
\beta_{IJ}=i\int dr\;
\;F_I^{\rm in}
\left(
\stackrel{\leftrightarrow}{\partial}_t
-\sigma f 
\stackrel{\leftrightarrow}{\partial}_r
\right)
F_J^{\rm out}
\,,
\end{eqnarray}
with the quantum numbers $I=(\xi,\omega)$ and
$J=(\xi',\omega')$. 
Inserting the result of the previous Section 
${\partial}_t S_\xi-\sigma f {\partial}_r S_\xi=\xi{\partial}_r S_\xi$
we arrive at
\begin{eqnarray}
\label{landau}
\beta_{IJ}
&=&
\int dr\,
A_{\xi}
\exp\left\{-i\omega S_\xi\right\}\,
\nonumber\\
&&\times\;
\frac{i\omega\xi\partial_rS_\xi-i\omega'(1+\sigma f [\sigma-\xi']/\chi)}
{\sqrt{\omega\omega'}}\,
\chi^{i(\sigma-\xi')\omega' R}
\nonumber\\
&&\times\;
{\cal N}
\left(1+{\cal O}[\chi]\right)
\left(1+{\cal O}\left[\frac{1}{\omega}\right]\right)
\,.
\end{eqnarray}
At this stage the correct meaning of the Landau symbols 
${\cal O}[\chi]$ and ${\cal O}[1/\omega]$ should be explained:
All terms which are of higher order in $\chi$ {\em and} of the same
order in $\omega$ as well as all terms which are of lower order in
$\omega$ {\em and} of the same order in $\chi$ -- in comparison with
the leading contributions in the integrand above -- are neglected.  
Such a detailed consideration is especially necessary for the quantity 
$\exp\{-i\omega S_\xi\}$ which involves terms like $(\chi\omega)^n$.
These contributions are {\em not} neglected -- in contrast to terms like
$\chi(\chi\omega)^n$.    

Accordingly, (exploiting the dominance of the vicinity of the horizon)
we may Taylor expand the amplitude  
$A_{\xi,\omega}(t=0,r)=A_{\xi,\omega}(t=0,r=R)+{\cal O}[\chi]$.
The zeroth-order term can be absorbed into the overall normalization
factor $\cal N$ via
\begin{eqnarray}
{\cal N}\rightarrow{\cal N}A_{\xi,\omega}(t=0,r=R)
\,,
\end{eqnarray}
and the higher order terms are omitted. 
(Here and in the following we do not change the symbol $\cal N$ for
the normalization factor and use the same letter also for the
modified  pre-factors.)
A similar procedure can be performed with the phase function $S_\xi$.
But owing to the pre-factor $\omega$ it is necessary to expand it up
to first order
\begin{eqnarray}
S_\xi(r)=
S_\xi(r=R)+\partial_rS_\xi(r=R)R\chi+{\cal O}[\chi^2]\,,
\end{eqnarray}
cf.~the remarks after Eq.~(\ref{landau}). 
Again the zeroth-order term $S_\xi(t=0,r=R)$ may be absorbed by a
redefinition of $\cal N$ 
\begin{eqnarray}
{\cal N}\rightarrow{\cal N}\exp\left\{-i\omega S_\xi(t=0,r=R)\right\}
\,.
\end{eqnarray}
Since we have to integrate over the initial frequency $\omega$ in 
Eq.~(\ref{w-int}) in order to obtain the number of created particles, 
the remaining unknown first-order term
$\partial_rS_\xi(t=0,r=R)$ can be eliminated by a re-scaling of the
initial frequency
\begin{eqnarray}
\omega\rightarrow\tilde\omega=\omega\,\xi\,\partial_rS_\xi(t=0,r=R)
\,.
\end{eqnarray}
Of course, such a transformation may be accomplished if and only if 
$\xi\partial_r S_\xi$ is positive.
But according to the arguments at the end of the previous Section 
the sign of $\partial_r S_\xi$ does not change during the
collapse -- as long as it occurs fast enough and regularly.
In this situation the four different branches of Eq.~(\ref{branch}) do
not mix and thus the sign of $\partial_r S_\xi$ equals its
initial value, i.e.~$\xi$.
Again we may consider a very abrupt change of the metric 
(sudden approximation, see the previous Section)
as an illustrative example, where the final phase function nearly
coincides with its initial form. 
For the Minkowski example it is simply determined by 
$\partial_rS_\xi\approx\xi$ and no redefinition is necessary
at all. For other initial metrics the redefinition
of the frequency exactly corresponds to the fact that the Hawking
effect is independent of the initial (regular and stationary)
space-time.
    
The Jacobi factor arising from the change of the $\omega$-integral
measure in Eq.~(\ref{w-int}) again modifies the normalization $\cal N$
only.  
This undetermined normalization factor will be fixed later
by virtue of the completeness relation in Eq.~(\ref{unit}) below.
After an analogous Taylor expansion of the function 
$f(t=0,r)=1+{\cal O}[\chi]$ we find 
\begin{eqnarray}
\label{betaR}
\beta_{IJ}
&=&
\int\limits_0^\infty d\chi\,
\exp\left\{-i\xi\tilde\omega R\chi\right\}\,
\frac{\tilde\omega-\omega'\sigma[\sigma-\xi']/\chi}
{\sqrt{\tilde\omega\omega'}}\,
\nonumber\\
&&\times\;
\chi^{i(\sigma-\xi')\omega' R}\,
{\cal N}
\left(1+{\cal O}[\chi]\right)
\left(1+{\cal O}\left[\frac{1}{\omega}\right]\right)
\,.
\end{eqnarray}
As expected from the previous considerations, the Bogoliubov
$\beta$-coefficients contribute only for $\sigma=-\xi'$ and vanish
(in leading order) for $\sigma=\xi'$.
In that case the out-modes are not singular (at the horizon)
-- only for $\sigma=-\xi'$ they display the arbitrarily fast
oscillating behavior.
Hence -- depending on the sign $\sigma$ -- either only ingoing 
(for $\sigma=-1$ and thus $\xi'=+1$) or only
outgoing 
(for $\sigma=+1$ and thus $\xi'=-1$)
particles are produced (in an infinite amount).  

The integral in Eq.~(\ref{betaR}) involves generalized {{\em eigen\,}}functions
which do not belong to the Hilbert space $L_2$ but are distributions,
cf.~\cite{blackholes}. Hence it cannot be interpreted as a
well-defined Riemann integral. But -- as demonstrated in
Ref.~\cite{blackholes} -- it is possible to approximate (locally)
the generalized {{\em eigen\,}}functions by well-defined wave-packets.
One way to simulate such an approximation is to introduce a
convergence factor via
$\chi^\varepsilon\,\exp\{-\varepsilon\chi\}$
with $\varepsilon\downarrow0$

For $\sigma=-\xi'$ the above integral can be solved in terms of 
$\Gamma$-functions.
After insertion of the convergence factor we can make use of the
formula \cite{abra}
\begin{eqnarray}
\label{gamma}
\int\limits_0^\infty dx\,e^{-xy}\;x^{z-1}=y^{-z}\,\Gamma(z)
\,,
\end{eqnarray}
which holds for $\Re(y)>0$ and $\Re(z)>0$, 
and -- remembering $\Gamma(z+1)=z\,\Gamma(z)$ -- we arrive at
\begin{eqnarray}
\label{gammabeta}
\beta_{IJ}
&=&
{\cal N}\,
\delta_{\sigma,-\xi'}\,\delta_{\xi,\xi'}\,
\sqrt{\frac{\omega'}{\tilde\omega}}\;\Gamma(2i\sigma\omega'R)\;
\left(i\xi\tilde\omega R+\varepsilon\right)^{2i\xi'\omega'R}
\nonumber\\
&&\times
\left(1+{\cal O}\left[\frac{1}{\omega}\right]\right)
\,.
\end{eqnarray}
In view of Eq.~(\ref{gamma}) the higher order terms in $\chi$ 
-- i.e.~$x$ -- cause increasing arguments $z$. 
Ergo these terms result in higher orders in
$1/y$ -- i.e.~$1/\omega$ -- consistently with our approximation and
the arguments at the beginning of this Section.
In order to evaluate the absolute value squared of the
$\beta$-coefficient we may utilize the identity \cite{abra} 
\begin{eqnarray}
\Gamma(z)\Gamma(-z)=-\frac{\pi}{z\sin\pi z}
\end{eqnarray}
to obtain the final result
\begin{eqnarray}
\label{beta^2}
\left|\beta_{IJ}\right|^2=
\frac{{\cal N}}{\tilde\omega}\,
\frac
{\delta_{\sigma,-\xi'}\,\delta_{\xi,\xi'}}
{\exp\{4\pi\omega'R\}-1}
\left(1+{\cal O}\left[\frac{1}{\omega}\right]\right)
\,.
\end{eqnarray}
This expression confirms the argumentation at the beginning of this
Section. The remaining integration over $\omega$ (or $\tilde\omega$) 
is indeed UV-divergent.
In addition we observe that the terms of higher order in $1/\omega$
(and thus $\chi$) that we have neglected in our calculations are not
UV-divergent and hence do not contribute to the Hawking effect.
This observation provides an {\em a posteriori} justification of our
expansion into powers of $1/\omega$ and $\chi$ and the neglect of the
sub-leading contributions.

The UV-divergence can be interpreted with the aid of the well-known
completeness relation  
\begin{eqnarray}
\label{unit}
\sumint{I}\alpha_{IJ}\alpha_{IK}^*-
\beta_{IJ}\beta_{IK}^*=
\delta(J,K)
\,,
\end{eqnarray}
where $I$ again symbolizes the initial quantum number.
This equality reflects the completeness of the initial modes.
Special care is required concerning the derivation of an analogue
expression involving the out-modes since some of those solutions are
restricted to the region inside or outside the horizon, respectively,
and these restricted modes are not complete in the full space-time.
In order to apply this relation we have to deduce the
$\alpha$-coefficients as well. 
For that purpose we define slightly modified Bogoliubov coefficients
via 
\begin{eqnarray}
\breve\beta(\omega,\omega')
=
\sqrt{\omega\omega'}\beta_{\omega\omega'}
\,,
\end{eqnarray}
and in analogy the $\alpha$-coefficient.
The modified Bogoliubov coefficients can be analytically continued
into the complex $\omega'$-plane where the relations  
$\breve F^*_{\rm out}(\omega')=\breve F_{\rm out}(-\omega')$ and hence
\begin{eqnarray}
\breve\alpha(\omega,\omega')
=
\breve\beta(\omega,-\omega')
\end{eqnarray}
hold. This enables us to derive the Bogoliubov $\alpha$-coefficient
for large initial frequencies $\omega$. Substituting
$\omega'\rightarrow-\omega'$ in Eq.~(\ref{gammabeta}) together with the 
complex conjugation the only difference  between $|\alpha_{IJ}|$ and
$|\beta_{IJ}|$ is the sign in front of the term $i\xi\omega R$.
Dividing the absolute values of the two coefficients all other terms
cancel and the convergence factor $\varepsilon$ determines the side of
the branch cut of the logarithm in the complex plane.
Hence we find for large frequencies $\omega$
\begin{eqnarray}
\label{bexpa}
|\beta_{IJ}|=
\exp\{-2\pi\omega'R\}\,
|\alpha_{IJ}|
\left(1+{\cal O}\left[\frac{1}{\omega}\right]\right)
\,.
\end{eqnarray}
Inserting Eq.~(\ref{bexpa}) into the completeness relation
(\ref{unit}) and considering the (singular) coincidence
$J=K$ it follows
\begin{eqnarray}
\label{N}
N_{J}
&=&
\bra{0_{\rm in}}\hat N_J^{\rm out}\ket{0_{\rm in}}=
\sumint{I}
\left|\beta_{IJ}\right|^2
\nonumber\\
&=&
\delta_{\sigma,-\xi'}\,
\frac{\delta_-(I,I)}{\exp\{4\pi\omega'R\}-1}
+{\rm finite}
\nonumber\\
&=&
\delta_{\sigma,-\xi'}\,
\frac{{\cal N}^-_V V}{\exp\{4\pi\omega'R\}-1}
+{\rm finite}
\,.
\end{eqnarray}
According to the results of Ref.~\cite{blackholes} the UV-divergence
of the $\omega$-integration of the absolute values squared of the
$\beta$-coefficients in Eq.~(\ref{beta^2}) exactly corresponds to the 
singular quantity $\delta_-(I,I)=\delta_-(\omega,\omega)$ and thus
represents the near-horizon ($r\downarrow R$, i.e.~$r_*\downarrow-\infty$)
part ${\cal N}^-_V V$ of the infinite volume divergence 
${\cal N}_V V={\cal N}^-_V V+{\cal N}^+_V V$
of the continuum normalization.  
As explained in Ref.~\cite{blackholes}, the infinitely large amount of
particles is necessary for (quasi) thermal behavior in an unbounded
volume.

It is also possible to calculate the Bogoliubov coefficients for
regular modes (wave packets instead of plane waves),
cf.~\cite{blackholes}. In this case no divergences occur and all
quantities are finite. 
Thus the (late-time) Hawking effect cannot be
easily distinguished from the the 
(collapse-dependent or initially present) finite amount of particles
via isolating the divergent part in this situation. 

As mentioned before, an initial state $\varrho_{\rm in}$ containing a
finite number of particles does not change the final results
concerning the Hawking effect. 
Inserting the Bogoliubov transformation the expectation value counting the 
number of Schwarzschild particles equals the Hawking term plus additional 
contributions in this situation 
\begin{eqnarray}
\varrho_{\rm in}\left(\hat N^{\rm out}_{J}\right)
&=&
N^{\rm Hawking}_{J}
\nonumber\\
&&
+
\sumint{IK}
\left(\alpha^{*}_{JI}\alpha_{JK}
+\beta^{*}_{JI}\beta_{JK}\right)
\varrho_{\rm in}\left(\hat A^{\dagger}_{I}\hat A_{K}\right)
\nonumber\\
&&
+
\sumint{IK}
\alpha^{*}_{JI}\,\beta_{JK}\,
\varrho_{\rm in}\left(\hat A^{\dagger}_{I}\hat A^{\dagger}_{K}\right)
\nonumber\\
&&
+
\sumint{IK}
\beta^{*}_{JI}\,\alpha_{JK}\,
\varrho_{\rm in}\left(\hat A_{I}\hat A_{K}\right)
\,.
\end{eqnarray}
For a state $\varrho_{\rm in}$ that contains a finite number of
initial particles the above expectation values vanish in the high
(initial) frequency limit $\omega_I,\omega_K\uparrow\infty$.
As a result the $I$ and $K$ summations/integrations are not
UV-divergent. Hence the additional contributions are finite and do not 
affect the (divergent) Hawking effect. 
E.g., if we assume the collapsing object to be enclosed by a (arbitrarily
large but finite) box with Dirichlet boundary conditions we may
describe an initial thermal equilibrium state via the canonical
ensemble. In view of the previous arguments we arrive at the
conclusion that any initial temperature does also not affect the final 
(Hawking) temperature in this scenario.
(Dropping the assumption of a finite box enclosing the collapsing
object the 
situation becomes more complicated. In this case the number of particles
being present initially diverges owing to the infinite volume. Hence
the usual spatial infinity part of the infinite volume divergence still
corresponds to the initial temperature whereas the near-horizon part
obeys the Hawking temperature, cf.~\cite{blackholes}.)
 
With the aid of similar arguments one can show that the Hawking effect
-- i.e.~the late-time radiation -- is also independent of the initial
metric (as long as it is regular). 
The number of particles created during the transition from one to
another regular metric is finite. These particles disperse after some
finite period of time and do not affect the (divergent) late-time part
of the radiation in accordance with the arguments in the previous
paragraph. 
In terms of the Bogoliubov coefficients this degree of freedom exactly
corresponds to the redefinition of the initial frequency $\omega$. 
(We did not need to specify the initial metric $f_{\rm in}(r)$ in
Sec.~\ref{PGL}.)

\subsection{Energy-momentum tensor}\label{energy}

In order to support the conclusions of the previous sections
concerning the evaporation/anti-evaporation we calculate the late-time 
expectation value of the relevant component of the energy-momentum
tensor. In contrast to the inherently non-covariant particle concept
this quantity is manifestly covariant.  
The general relativistic energy-momentum tensor 
$T_{\mu\nu}$ as the source term in
the Einstein equations can be obtained via variation of the action
$\cal A$
with respect to the metric. For the scalar field under consideration
we obtain the well-known expression 
\begin{eqnarray}
\label{Tmunu}
T_{\mu\nu}=\frac{2}{\sqrt{-g}}\,\frac{\delta\cal A}{\delta g^{\mu\nu}}=
\partial_\mu\Phi\,\partial_\nu\Phi-\frac{g_{\mu\nu}}{2}\,
\partial_\gamma\Phi\,\partial^\gamma\Phi
\,.
\end{eqnarray}
The covariant divergence of this tensor vanishes
\begin{eqnarray}
\label{covdiv}
\nabla_\mu\,T^\mu_\nu=\frac{1}{\sqrt{-g}}\;
\partial_\mu\left(\sqrt{-g}\;T^\mu_\nu\right)-
\frac{1}{2}\;T^{\alpha\beta}\,\partial_\nu\,g_{\alpha\beta}=0
\,.
\end{eqnarray}
However, in general this equality does not imply any conservation law
due to the exchange of energy and momentum between the scalar and the 
gravitational field (second term).
Nevertheless, if the space-time possesses a Killing vector $\zeta_\nu$ 
mediating the time-translation symmetry (Noether theorem) we may
define a conserved energy current 
\begin{eqnarray}
J^\mu=T^{\mu\nu}\,\zeta_\nu
\,.
\end{eqnarray}
In view of the symmetry of the energy-momentum tensor 
$T^{\mu\nu}=T^{\nu\mu}$ and Eq.~(\ref{covdiv}) together with the
property of the Killing vector $\nabla_\mu\zeta_\nu+\nabla_\nu\zeta_\mu=0$  
this current is indeed conserved
\begin{eqnarray}
\nabla_\mu J^\mu=0
\,.
\end{eqnarray}
Now we may calculate the energy flux $\Xi$ out of (or into) the black 
(white) hole
\begin{eqnarray}
\Xi
=
\int\limits_\Sigma d\Sigma_\mu\,J^\mu
=
\int d\Sigma_\mu\,
\bra{0_{\rm in}}\hat T^{\mu\nu}\ket{0_{\rm in}}\,\zeta_\nu 
\,,
\end{eqnarray}
where $\Sigma$ denotes the (cylindrical) hyper-surface enclosing
the black/white hole. 
In a 3+1 dimensional space-time one may determine $\Sigma$ via the
Killing vectors mediating the spherical symmetry.
By virtue of the Gauss law the above quantity is invariant under
deformations of this hyper-surface $\Sigma$. Hence we may consider
a sphere with a radius which is much larger than the Schwarzschild
radius -- where the metric coincides asymptotically
($r\uparrow\infty$) with the Minkowski form. 
In this region the energy flux simplifies to 
\begin{eqnarray}
\label{flux}
\Xi
=
-\frac12
\int dt\,\bra{0_{\rm in}}\left\{
\frac{\partial\hat\Phi}{\partial t},
\frac{\partial\hat\Phi}{\partial r}
\right\}\ket{0_{\rm in}}
\,.
\end{eqnarray}
The symmetrization $\{\cdot,\cdot\}$ is necessary in order to obtain a
Hermitian observable $\hat T^{\mu\nu}$. The minus sign arises from 
$g_{11}(r\uparrow\infty)=-1$.
For large radial distances (approximately Minkowski) the expansion of
the field reads  
\begin{eqnarray}
\hat\Phi(t,r)
&=&
\sumint{I}
\hat a_I^{\rm out}\,F_I^{\rm out}(t,r)+{\rm h.c.}
\nonumber\\
&=&
\sumint{\omega\xi}
\frac{\cal N}{\sqrt{\omega}}\,\hat a^{\rm out}_{\omega\xi}
\,e^{-i\omega(t+\xi r)}
+{\rm h.c.}
\end{eqnarray}
Insertion of the above expansion into the bilinear form in
Eq.~(\ref{flux}) generates a 
sum over $\xi$ and $\xi'$ as well as an integration over $\omega$ and  
$\omega'$. The time-integration in Eq.~(\ref{flux}) involves terms
such as $\exp\{\pm i \omega t \pm i \omega' t\}$ and thus generates 
$\delta(\omega\pm\omega')$-distributions. In view of the positivity
of the frequencies only $\omega=\omega'$ contributes. 
Similarly the remaining spatial dependence 
$\exp\{\pm i \omega (\xi-\xi') r\}$ implies that merely $\xi=\xi'$
yields relevant contributions at large distances $r\uparrow\infty$.
As a result only one $(\omega,\xi)$-summation/integration survives and 
the late-time radiation is related to the number of particles via   
\begin{eqnarray}
\Xi=
-|{\cal N}|^2
\sumint{\omega\xi}
\bra{0_{\rm in}}
\hat N^{\rm out}_{\omega\xi}
\ket{0_{\rm in}}
\omega\,\xi
\,.
\end{eqnarray}
This relation confirms the conclusions of the previous sections: 
The divergence of 
$\bra{0_{\rm in}}\hat N^{\rm out}_{\omega\xi}\ket{0_{\rm in}}$
exactly corresponds
to the time-integration and the resulting singularity of the 
$\delta(\omega-\omega')$-distribution.
The Bogoliubov coefficients and thus also 
$\bra{0_{\rm in}}\hat N^{\rm out}_{\omega\xi}\ket{0_{\rm in}}$
contribute (in an infinite amount) only for $\sigma=-\xi$.
Hence the collapse to a black hole described by the branch 
$\sigma=+1$ of the PGL metric generates an outward ($\xi=-1$) flux at
late times whereas the collapse to a white hole corresponding to
$\sigma=-1$ leads to an inward ($\xi=+1$) flux at late times.

\section{Summary}\label{summary}

In terms of the Painlev{\'e}-Gullstrand-Lema{\^\i}tre coordinates it
is possible to model a gravitational collapse of an object and the
subsequent formation of an apparent horizon by means of a
manifestly $C^\infty$-metric. This set of coordinates possesses two
separate branches (labeled by $\sigma=\pm1$). Depending on the
particular branch (i.e.~the sign of $\sigma$) either only ingoing or
only outgoing particles are created in an infinite amount. 
This infinite amount of particles obeys a thermal spectrum
corresponding to the Hawking temperature.

\section{Conclusions}\label{conclusions}

The theorems presented in Section \ref{hadamard} imply that during
every collapse scenario that can be described by a $C^\infty$-metric
an infinite number of particles with a thermal spectrum corresponding
to the Hawking temperature is created.
This statement is verified in the present article for a rather general
{\em ansatz} for a $C^\infty$-metric in Eq.~(\ref{dynPGL}). 
For that purpose it is neither necessary to impose any conditions on
the metric beyond the horizon nor to specify the explicit dynamics of 
$f(t,r)$ during the collapse -- 
as long as it is regular, i.e.~$C^\infty$, and fast enough, 
cf.~Sec.~\ref{eikonal-ansatz}.

So the Hawking effect is not the result of a space-time
singularity but a consequence of the formation of a horizon
(strictly speaking, an apparent horizon).   
For the derivation of the Bogoliubov coefficients no assertions about  
the metric in the interior region $f(t\geq0,r<R)$ are necessary at 
all. 
In addition, only the modes that are affected by the horizon 
(one-way membrane, cf.~\cite{visser}) contribute to the late-time 
(Hawking) radiation, i.e.~the outgoing particles for the black hole
horizon and the ingoing particles for the withe hole horizon,
respectively.   

Thus the properties of the produced particles crucially depend on
the branch of the Painlev{\'e}-Gullstrand-Lema{\^\i}tre metric under
consideration. 
Adopting the Schr{\"o}dinger representation the two distinct branches
generate completely different final states $\varrho_\sigma$.
Only one state represents the phenomenon of evaporation
while the other state corresponds to anti-evaporation.

\section{Discussion}\label{discussion}

Perhaps the most striking outcome of the presented calculation is the
fact that -- depending on the particular branch $\sigma$ of the
dynamics during the collapse -- the final state of the quantum field
does not necessarily represent evaporation but possibly also
anti-evaporation. 
The phenomenon of anti-evaporation has already been discussed in the
literature, see e.g.~\cite{bousso,nojiri}, but in a different
context 
(Schwarzschild-de Sitter geometries, see also \cite{hiscock}). 
In contrast the calculation in the present article applies to
asymptotically flat space-times.  

For one branch the final state coincides -- up to a finite number of
particles -- with the 1+1 dimensional analogue of the Unruh state
$\varrho_{\rm U}$ describing evaporation. 
The other branch generates the (in some sense) opposite final state 
-- corresponding to anti-evaporation.  
In the following considerations we shall denote this state 
as the anti-Unruh state $\varrho_{\rm aU}$ for convenience. 
This state $\varrho_{\rm aU}$ can be obtained from the Unruh state
$\varrho_{\rm U}$ by means of the (Schwarzschild) 
time-inversion ${\sf T}$ 
\begin{eqnarray}
\varrho_{\rm aU}={\sf T}\varrho_{\rm U}
\,,
\end{eqnarray}
if we regard the (Schwarzschild) metric of the space-time as fixed.
Induced by the time-inversion ${\sf T}$ all outgoing particles turn
their direction into ingoing and {\em vice versa}.
Since the neutral scalar field is neither affected by the charge
conjugation ${\sf C}$ nor by the parity transformation ${\sf P}$ 
(in contrast to a pseudo-scalar field) and we consider a
spherically symmetric situation, both, the Unruh as well as the
anti-Unruh state are not ${\sf CPT}$ invariant: 
${\sf CPT}\varrho_{\rm U}=\varrho_{\rm aU}\neq\varrho_{\rm U}$. 
These considerations are relevant for the investigation of unitarity
and time-reversibility, see e.g.~\cite{strominger,rama}.

Searching for the physical implementations of the main result of the
present article there are several possible interpretations:

From a conservative point of view one might argue that 
{\em the branch causing anti-evaporation is unphysical} and should be
excluded. 
This assertion might perhaps be supported by physically reasonable
constraints on the energy-momentum tensor, such as the energy
conditions.  
The two branches of the $C^\infty$-metric in Eq.~(\ref{dynPGL}) 
-- after the straightforward generalization to 3+1 dimensions --
can be used to derive the associated Ricci tensor $R_{\mu\nu}$.
Owing to the smoothness of the metric the curvature tensor always
exists and is $C^\infty$ as well.
By virtue of Einstein's equations the Ricci tensor reveals the  
corresponding energy-momentum tensor which could be 
compared with an appropriate model of a collapsing star or 
used to test the energy conditions, for example.
It is well-known that appropriate energy conditions exclude the
existence of some pathological space-time scenarios, such as
certain worm-holes or time-machines, see e.g.~\cite{visser}.
However, one should be aware that the energy conditions may well be
violated if one incorporates the back-reaction of the quantum field. 
Since the Hawking effect is most relevant for small objects and almost
negligible at astrophysical orders of magnitude one would expect that
such quantum effects have to be taken into account.  

As another possible interpretation of the result of this article one
may arrive at the conclusion that 
{\em black holes evaporate but white holes anti-evaporate}.
The particle production by white holes has already been discussed in 
Ref.~\cite{rama}, but within a different context:
The space-time under consideration in Ref.~\cite{rama} was obtained
via the time-inversion of a space-time representing the collapse of an
object to a black hole, i.e.~an anti-collapse.
In contrast the space-time investigated in the present article
corresponds to the gravitational collapse of an object,
cf.~Figs.~\ref{fig4} and \ref{fig5}.   
Furthermore the initial state in Ref.~\cite{rama} is determined via a
factorization assumption which might be questioned in general and does 
definitely not apply to the scenario of the present article.
As a consequence the resulting radiation becomes singular at the
retarded time of the termination of the horizon -- a prediction which 
differs drastically from the outcome of the present article. 
Based on similar scenarios in Refs.~\cite{zeldovich} and \cite{eardley} 
further instabilities and quantum effects connected with white holes
are discussed -- before (and independently of) Hawking's discovery. 
The comparison of black and white holes is potentially interesting
in view of the fundamental question of time-reversibility 
(unitarity and the second law of thermodynamics) of quantum gravity,
see e.g.~\cite{rama}.  

The predicted anti-evaporation of white holes is certainly relevant for
the sonic black/withe hole analogues, see 
e.g.~\cite{sonic,acoustic,volovik,fischer}.
These flow profiles always possess an (effective) black hole {\em and} 
a white hole horizon, see e.g.~Fig.~1 in Ref.~\cite{volovik}. 
If the fluid accelerates in such a way that its local velocity exceeds
the speed of the sound (black hole horizon) it decelerates below
the speed of the sound somewhere (white hole horizon) as well.
Consequently, the derivation presented in this article implies that, 
if the perturbations of the flow profile of the liquid obey a quantum
field theoretical description (with the resulting quantum fluctuations),
then the associated (effective) vacuum fluctuations are converted into 
(quasi) particles leading to evaporation for the black hole horizon
and to anti-evaporation for the white hole horizon. 

In order to discuss the third opportunity regarding the interpretation
of the outcome of the present article one may recall the fact that
the Hawking effect is dominated by arbitrarily large (initial)
frequencies. 
But at energies above the Planck scale one expects the breakdown of
the treatment of quantum fields propagating in given 
(externally prescribed) space-times. In the Planck {\em r\'egime} 
the back-reaction, for example, should become important.
Assertions about the metric in this region 
(e.g.~the Planck scale vicinity of the forming horizon)
are a very delicate issue. 
Hence one is lead to the assumption that 
{\em the outside observer cannot distinguish the two branches}.
As any explorer at a finite spatial distance to the collapsing object 
merely experiences the static Schwarzschild metric (Birkhoff theorem), 
the only possible way to obtain informations about the collapse is
provided by the quantum radiation itself.   
If we now assume that the outside beholder cannot resolve the 
behavior of the metric in the Planck scale vicinity of the forming
horizon, he/she can obtain no information about the particular branch
of the metric {\em a priori}.
Without any knowledge about the value of $\sigma$ during the
collapse the most natural {\em ansatz} for the state governing the
measurements of an outside observer is given by -- remember the
convexity of the states discussed in Sec.~\ref{formalism} 
\begin{eqnarray}
\varrho_0=\frac{\varrho_{\rm U}+\varrho_{\rm aU}}{2}
\,.
\end{eqnarray}
Again we adopt the Schr{\"o}dinger representation.
This {\em ansatz} complies with the superposition principle of quantum
theory -- if one assumes that the back-reaction of the quantum fields
onto the metric yields relevant contributions.

The above introduced state describes some kind of quasi-thermal
equilibrium -- it 
contains the same (infinite) number of ingoing and outgoing particles
with a thermal spectrum corresponding to the Hawking temperature.
Although the state $\varrho_0$ displays in 1+1 dimensions a close
similarity to the (1+1 dimensional analogue of the)
Israel-Hartle-Hawking state $\varrho_{\rm IHH}$,
in 3+1 dimensions this quasi-thermal equilibrium state $\varrho_0$
differs drastically from the Israel-Hartle-Hawking state, which
describes (at least with respect to the algebra of observables outside
the horizon) real thermal equilibrium. 
The expectation value of the number
of particles in the Israel-Hartle-Hawking state $\varrho_{\rm IHH}$
exhibits the complete infinite volume divergence, i.e.~the
near-horizon part $r_*\downarrow-\infty$ as well as the usual spatial
infinity $r_*\uparrow\infty$, cf.~\cite{blackholes}.
In contrast the analogue expectation value in the states 
$\varrho_0$, $\varrho_{\rm U}$, and $\varrho_{\rm aU}$
contains the near-horizon part only, see Sec.~\ref{bogoliubov}.
As a consequence the renormalized expectation value of the energy
density in the states 
$\varrho_0$, $\varrho_{\rm U}$, and $\varrho_{\rm aU}$ 
decreases for large distances $r$ with $1/r^2$
whereas the same quantity approaches a constant value 
(in view of the Stefan-Boltzmann law proportional to $T^4$) 
in the Israel-Hartle-Hawking state $\varrho_{\rm IHH}$.   

It might be noted here that -- in contrast to the Unruh as well as the
anti-Unruh state -- the state $\varrho_0$ is ${\sf CPT}$ invariant: 
${\sf CPT}\varrho_0=\varrho_0$. 
Therefore the unitarity and time-reversibility problem mentioned above
in connection with the (anti) Unruh state does not necessarily apply
to this state. 
        
In Sec.~\ref{bogoliubov} we have observed that only the region near
the horizon 
generates contributions that are relevant with respect to the Hawking
effect. Exactly the leading terms in $1/\omega$ and $\chi$ give rise
to the UV-divergence accounting for the Hawking effect.
The notion of the vicinity of the horizon as the region that is
essential for the Hawking effect may be illustrated via the following  
{\em gedanken} experiment:
Let us imagine a very thin shell of matter with slowly decreasing radius.
As long as the radius of the shell is larger than the associated 
Schwarzschild radius the number of created and radiated particles 
remains finite as a consequence of the regularity of the metric and
the associated {{\em eigen\,}}modes.
If the shell were to stop shrinking before it reached its
Schwarzschild radius, no Hawking effect would be observed.
Accordingly, the creation of particles accounting for the Hawking effect
occurs exactly in the space-time region of the formation of the horizon.

In order to support the argumentation in Sec.~\ref{eikonal-ansatz} 
concerning the independence and separation of the different branches  
(e.g.~corresponding to ingoing and outgoing components)
in Eq.~(\ref{branch})
we may consider a conceptual clear scenario -- where the
effective boundary condition at $r=0$ does not contribute at all -- 
described in the following {\em gedanken} experiment: 
At first we suppose a small amount of highly charged matter to
collapse at the center of gravity forming a tiny extreme
Reissner black hole. The surface gravity of such a black hole
vanishes with the 
result that there is no Hawking radiation (at this stage).
After the formation of the small black hole the point $r=0$ is hidden
by the corresponding horizon. Consequently, there is no "reflection"
at the origin $r=0$ in this case.
(It is possible to define ingoing and outgoing particles separately,
cf.~\cite{blackholes}.)
If we now suppose the matter (enclosing the tiny black hole) to
collapse  
the origin cannot generate a mixing of the different branches 
(e.g.~ingoing and outgoing).

\section{Outlook}\label{outlook}

For the calculation of the Bogoliubov coefficients we have restricted
our consideration to the 1+1 dimensional space-time of the
$(t,r)$-sector. 
Even though the main result of this article should persist also in 3+1 
dimensions (including the angular terms), some additional
complications arise \cite{forthcoming}. 
The Klein-Fock-Gordon equation (\ref{KFG}) assumes for the 3+1
dimensional Schwarzschild metric a slightly modified form. 
After separating the angular dependence by spherical harmonics the
centrifugal barrier and curvature scattering effects can be
incorporated into an effective potential ${\cal V}_{\rm eff}$,
see e.g.~\cite{fredenhagen}
\begin{eqnarray}
\left(
\frac{\partial^2}{\partial t^2}
-\frac{\partial^2}{\partial r_*^2} 
+{\cal V}_{\rm eff}(r_*,\ell)
\right)\phi_{\ell,m}=0
\,.
\end{eqnarray}
${\cal V}_{\rm eff}$ is strictly positive and approaches zero for
$r_*\uparrow+\infty$ and for $r_*\downarrow-\infty$
with ${\cal O}[1/r^2]={\cal O}[1/r_*^2]$ and  
${\cal O}[\chi]={\cal O}[\exp\{r_*/R\}]$, respectively. 
Unfortunately, in 3+1 dimensions
no closed expression (in terms of well-known functions)
for the {{\em eigen\,}}modes is available. The asymptotic behavior can be
derived easily. For $r_*\downarrow-\infty$ the positive frequency
solutions again behave as $\exp\{-i\omega t\pm i\omega r_*\}$ or
linear combinations of them.
These waves are purely ingoing or outgoing, respectively, for
$r_*\downarrow-\infty$. 
But every mode which is purely outgoing near the horizon contains for
$r_*\uparrow+\infty$ ingoing components as well owing to the
scattering at the effective potential ${\cal V}_{\rm eff}$ 
(inducing transmission and reflection coefficients) and {\em vice versa}. 
In order to obtain a complete and orthogonal (with respect to the
inner product) set of positive frequency solutions of the 
Klein-Fock-Gordon equation one has to combine the different
opportunities. 
E.g., within the notation of Ref.~\cite{misner} (see Fig.~1 there) a
'down' mode is purely ingoing at infinity and (therefore) mixed at the 
horizon.  
According to the results of this article one would expect that an
infinite amount of particles are created by the collapse in this mode
-- independently of the branch of the metric \cite{forthcoming}.

Furthermore the present article considers the most simple example of a
quantum 
field theory, i.e.~the neutral, massless, and minimally coupled scalar 
field $\Phi$. Further investigations should be devoted to fields
obeying more complicated equations of motion. 
For the spin-zero field example one may incorporate potential terms
including masses $M^2\Phi^2$ or conformal couplings
$R^\mu_\mu\Phi^2/6$ and consider charged (i.e.~non-Hermitian) fields.
Moreover, it would be interesting to extend the examination to
fields with higher spin, e.g.~the electromagnetic field.
Nevertheless, there is no obvious reason why the main conclusions of
this article should not persist. 
The evaluation of the Hawking effect for interacting fields with
non-linear equations of motion seems to be rather challenging.

Similarly the space-time under consideration describes the most simple
example of a black hole. The Schwarzschild geometry represents an
uncharged and non-rotating black hole where the Einstein tensor and
thereby also the energy-momentum tensor vanish for $r>0$.
The extension of the results presented in this article to more general
static (i.e.~non-rotating) black-holes -- e.g.~the Reissner solution --
seems to be straight-forward, see also \cite{blackholes}.
In contrast the investigation of rotating (i.e.~stationary, but not static)
black-hole space-times -- e.g.~the Kerr solution -- holds more
difficulties.  

Apart from the Painlev{\'e}-Gullstrand-Lema{\^\i}tre coordinates there
are several other coordinate sets that describe the Schwarzschild
geometry space-time by a manifestly $C^\infty$-metric, e.g.~the
Eddington-Finkelstein coordinates.  
It might be interesting to consider a collapse model in terms of these
coordinates in analogy to Eq.~(\ref{dynPGL}) and to compare the results.   
 
However, one should be aware that all of the previous considerations
neglect the back-reaction of the quantum field onto the metric.
So far the quantum field is treated as a test field propagating on a
given (externally prescribed) space-time.
If one attempts to leave this formalism several problems arise:  
The concept of Hadamard states as described in Eq.~(\ref{hadi}) is
restricted to free fields obeying linear equations of motion.  
The two-point function of interacting fields possesses additional
singularities in general. Consequently -- if one regards the treatment
of quantum fields in classical (general relativistic) space-times as a
low-energy effective theory of some underlying theory -- the
imposition of the Hadamard condition is not obviously justified.
Similarly the requirement of a smooth $C^\infty$-metric may be
questioned from this point of view.
Accordingly, it might be interesting to examine the consequences of
collapse dynamics that are not $C^\infty$ regarding the Hawking
effect \cite{forthcoming}.

An exhaustive clarification of these problems probably requires the
knowledge of an underlying law that unifies quantum field theory and
general relativity. 


\section*{Acknowledgment}

The author is indebted to
A.~Calogeracos, K.~Fredenhagen, I.~B.~Khriplovich, G.~Plunien,
G.~Schaller, G.~Soff, and W.~G.~Unruh
for valuable discussions.
This work was partially supported by BMBF, DFG, and GSI. 

\addcontentsline{toc}{section}{References} 


\begin{thebibliography}{99}
\bibitem{forthcoming}
These considerations will be established in detail in a forthcoming
publication.
\end{thebibliography}
\end{document}